\documentclass[10pt]{article}
\usepackage[dvipdfm]{graphicx}
\begin{document}
\title{Mesoscale Quantization and Self-Organized Stability}
\author{Randall D. Peters\\
 Mercer University Physics}
\date{June 2005}
\maketitle
\bibliographystyle{plain}

\begin{abstract}
In the world of technology, one of the most important forms of friction
is that of rolling friction.  Yet it is one of the
least studied of all the known forms of energy dissipation.
In the present experiments we investigate
the oscillatory free-decay of a
rigid cube, whose side-length is less than the diameter of the rigid
cylinder on which it rests.
The resulting free-decay is one of harmonic motion with damping.
The non-dissipative character of the oscillation yields
to a linear differential equation;
however, the damping is found to involve more than a deterministic
nonlinearity.
Dominated by rolling friction, the damping is sensitive to the material
properties of the contact surfaces.  For `clean' surfaces of glass
on glass, the decay shows features of mesoscale quantization and
self-organized stability. 
\end{abstract}  

\section{Geometry}
Shown in cross-section in figure 1 are the two
objects that comprise this oscillatory
system.  The cube side-length dimension is $2a$, and the diameter of
the cylinder is $2r$.  Both are assumed to be completely rigid, so that
the torque which restores the cube toward equilibrium is due to gravity.
In mechanics
textbooks this problem is commonly treated as an exercise
to test for the condition of static equilibrium.  

\section{Theory without damping}
The potential energy is simply $mgh$.  From the other identities provided
in the figure, one can readily show for small displacements, that
\begin{equation}
U~=~\frac{mg}{2}(r-a)\theta^2
\end{equation}

The displacement of the center of mass P of the cube is given by
$\Delta x~=~r~sin~\theta~-~s~sin(\phi-\theta),\newline
~\Delta y~=~
s~cos(\phi-\theta)~-~a$.  For small $\theta$ these yield for the
center of mass velocity
\begin{equation}
v_x~\rightarrow~a\frac{d\theta}{dt}~=~a\dot{\theta},~~~~~~~~ v_y~\rightarrow~0
\end{equation}

The moment of inertia of the cube about its center of mass is given by
$I~=~\frac{2}{3}ma^2$.  The kinetic energy is the sum of translational
and rotational parts; i.e., $T~=~\frac{1}{2}mv^2+\frac{1}{2}I\dot{\theta}^2
~=~\frac{5}{6}ma^2\dot{\theta}^2$.  Using $p~=~\frac{\partial T}{\
\partial \dot{\theta}}$
the Hamiltonian is calculated to be
\begin{equation}
H(p,\theta)~=~\frac{3p^2}{10~ma^2}~+~\frac{mg}{2}(r-a)~\theta^2
\end{equation}

Using Hamilton's canonical equations, $\frac{\partial H}{\partial p}
~=~\dot{\theta}$ and $\frac{\partial H}{\partial \theta}~=~-\dot{p}$

one obtains the equation of motion
\begin{equation}
\ddot{\theta} ~+~\frac{3g}{5a^2}(r-a)\theta~=~0
\end{equation}
from which the period is determined to be

\begin{equation}
T~=~2\pi\sqrt{\frac{5a^2}{3g(r-a)}}
\end{equation}

\section{Adding friction to the model}
The naive approach to the addition of damping would be to add to
to the left hand side of eq.(4)
the term $2\beta\dot{\theta}$, where $\beta$ is sometimes referred to
as the damping constant.  In reality, in dozens of experiments performed
by the present author on mechanical oscillators-- not a single case has
ever been
found for which 
it is proper to call $\beta$ a constant.  Even if the present
system were amenable to the overly-idealized viscous damping approximation
indicated by this term, $\beta$ is never constant and should
be referred to as the viscous damping {\em coefficient}. With fluids
it depends
on the density as well as the viscosity, and it is never completely
independent of frequency\cite{01}.  

\subsection{Nonlinear damping}
From the experimental results it will be seen that the damping is certainly
not a classical exponential decay of the type students are conditioned
from their textbooks to expect for virtually every oscillatory
system in free-decay.
A closer approximation to most of the cases (although inadequate) is `simple'
Coulomb damping; i.e.,
where one adds to eq.(4) a term that is proportional to $sgn(\dot{\theta})$.
The signum function causes the effective damping force to depend 
on the direction of the velocity but not its magnitude.  It is
a nonlinear damping term, and the quality factor Q decreases linearly
with time,
as opposed to being constant \cite{01}.

\section{Experimental Setup}
Shown in figure 2 is a photograph of the apparatus that was
employed for
this study.  The material of both cylinder and cube is glass, components
normally used for optics experiments. All surfaces are highly polished,
so that visible light is specularly reflected from them.
The dimensions are $r~=~4.46~cm, \&~ a~=~2.54~cm$. The thin
half-cylinder (length = 1.4 cm) is glued at its bottom to the aluminum
block on which
it rests.  The block is in turn glued to the wooden base to which the
sensor support `post' is attached with a screw.

Visible in the picture is a
thin rectangular piece of sheet copper that was superglued to one side of the
cube.  Passing between static electrodes, it is the moving member of
the capacitive sensor that was used to measure the displacement of the cube.
The sensor is the symmetric differential capacitive (SDC) device invented
and patented by the author\cite{02}. 

The interface to the computer is a Dataq Inc. DI-154 analog to digital
12-bit converter.  It was used instead of their \$25 10-bit DI-194 to
permit lower-level signals to be studied.  The 10-bit ADC would be
satisfactory for many of the cases considered, with the SDC electronics
set at maximum gain. 

\section{Experimental Results}
\subsection{Frequency}
As a test of eq(5), the period of oscillation was measured using
the Fast Fourier Transform (FFT) of free-decay records.  A
low-level case is shown in the right hand graph of figure 3. The
second harmonic is down 33 dB
(factor of 45) from the fundamental.  It is not known whether this
harmonic distortion results primarily from the sensor or if it is
a characteristic
of the cube/cylinder oscillator.  It should be noted that the decibel,
as used in all graphs (excepting figure 5), is according to the
Dataq convention; i.e.,
\begin{equation}
dB~=~20~log_{10}(32768~v/v_{fs})
\end{equation}
\noindent where the full-scale voltage is for the present ADC
not adjustable; i.e., it is fixed at $v_{fs}~=~10 ~V$. 
Of course, as with potential energies, only differences are important;
making the reference
inconsequential. To convert to the more common dB relative to 1 volt, one
only needs to subtract 70.3 from the Dataq dB value.

All spectra, both low-level and high-level, yielded the same 2.2 Hz
eigenfrequency.  With the dimensions indicated above, the theoretical
estimate of the frequency from eq (5) is $f~=~\frac{1}{T}~=~2.1~ Hz$. This
agreement to within
5\% is considered reasonable, since the `regularity' of the cylinder
is unknown.  Although the cube is polished flat to approximately
the wavelength
of visible light; variability in curvature
around the cylinder as a function of angle is expected to be more
significant.

It should be noted that the moment of
inertia of the copper strip, which is not factored into the
theory, will lower the frequency only slightly; since
its mass is less than 2\% the cube's mass of 329 g.

Although the compressibility of the surfaces has not been considered in a
quantitative manner,
this may be important for efforts directed at better than
5\% agreement between theory and experiment.  The weight of the cube
flattening the cylinder and cupping the cube at the contact point,
causes the frequency to increase. 

\subsection{Damping}

As can be seen from the compressed-time record of figure 3, the free-decay
is not exponential.  Moreover, it is not even symmetric; i.e., the
upper and lower turning points of the motion are not mirror-symmetric
about the time-axis.

At low levels of the motion, where the cube's resting position
on the cylinder can be
maintained nearly constant, repeatibility between runs was
found to be good, as seen in the example of figure 4. To initialize the
oscillation at these low levels, a stream of air was directed 
onto a side of the cube (a strong puff from the mouth).  At high levels
the oscillation was accomplished by placing a screwdriver under one side
of the wooden base, tilting it by a small angle and then removing
the tilt rapidly (but not by dropping).  

The nonlinearities observed in figure 4 derive from the oscillator and
not from the sensor.  The sensor was determined to be essentially linear
over the full operating range of plus and minus voltages considered
in the study.
The sensor was calibrated using an optical lever; i.e.,
by means of a laser pointer directed onto one surface of the cube (using
a screen with a meter stick placed 3.6 m away). The calibration constant
was measured to be 48 V/rad.\\

{\em Damping Theory}\\

Figure 5 illustrates the differences among hysteretic, Coulomb, and
amplitude dependent damping.  Hysteretic damping yields an exponential
free-decay and may be approximated (for fixed frequency) by the viscous
damping model.  In actuality, all three of the damping cases shown in
figure 5 are nonlinear.  Only the hysteretic case masquerades as
linear \cite{01}.

Comparing figures 4 and 5, one observes that the oscillating cube is
closest to the case of Coulomb damping.  Although one normally associates
Coulomb damping with sliding surfaces, we see that it is also a first
approximation to the rolling friction of the present study. Because the
turning points of the motion deviate from a line (or other simple curve)
a non-traditional method of measuring the damping was employed.

\subsection{Short Time Fourier Transform}
The left set of curves in figure 5 may be familiar to the reader, but
the middle set of curves is less common. 
Because the decay curves of this study are not simple, the method
selected for measuring the damping is one called the short time Fourier
transform (STFT).  The time-record is analyzed, using the Fast Fourier
transform (FFT), by evaluating at each of a set of points that are
equally spaced in time, and which cover the full span of the record.
For a given length record, the maximum allowable number of STFT points
is determined by the $\Delta t$
between points and the number of points used in calculating the FFT.
The value of $\Delta t$ is governed by the sample rate of the ADC
and the total duration of the decay.  For all present work the sample
rate was $240~s^{-1}$. For most of the graphs the time between FFT points
was 5.2 s, a value made convenient by the modus operandus of the Dataq
software. 
Also for the present work, the number of points chosen for each FFT
calculation
was 1024. Spectra calculated with anything smaller (including 512)
yield too poor resolution to be acceptable.
For high-level decays with small damping this permitted more
than 30 points to be plotted.  At low-levels the number was reduced by
a factor of about two for small damping cases.  When the damping was
large, due to surface contamination, the Q was so small as to disallow
an accurate assessment of the damping history and thus the
mechanism (too few points).
From a given spectrum, the level of the fundamental component
was read in decibels; this value was then plotted in time at
the midpoint of the 1024 point-spread.

As seen from figure 5 (middle graphs), such an STFT yields a straight
line if it was obtained from an exponential (hysteretic) decay.  If
the decay curve derives from Coulomb friction, the STFT curve falls
at an ever increasing
rate, becoming precipitous at the end time.  For amplitude dependent
damping (such as from drag of a fluid that is quadratic in the speed), the
opposite trend is found.

\subsection{Quality Factor}
The quality factor in free-decay is defined according to
\begin{equation}
Q~=~2\pi\frac{E}{|\Delta E|}
\end{equation}
\noindent where $E$ is the energy of the oscillator at a given time, and
$|\Delta E|$ is the energy loss to the damping friction per cycle at that
time.  For the case of hysteretic damping $Q$ is constant, as observed in
the right set of graphs of figure 5. The exponential decay requires
of the amplitude $A$ that
$\Delta A/A~=~constant$, giving $Q~=~constant$, since $E~\propto~A^2$.

The $Q$ of Coulomb damping decreases
linearly in time as shown in the graph.  This can be easily understood,
since the friction force is constant, and the work done against friction
is thus proportional to the amplitude.  Noting again that the energy is
proportional
to the square of the amplitude, the $Q$ for this case must then be
proportional to the amplitude, which itself falls off linearly.
To calculate the Q from the falloff slope in $dB/s$, one uses the
expression $Q~=~27.29\frac{f}{|dB/s|}$, where f is the frequency of
oscillation.

A thorough quantitative treatment of these nonlinear damping mechanisms,
including the amplitude dependent case,
can be found in \cite{01}.

\section{Damping Measurements}
Shown in figure 6 are graphs of STFT and Q vs time for the left case of
figure 4. The history of Q is remarkable in its non-classical properties.
There are three separate regions of
essentially constant slope, implying a different constant 
`coefficient' of
(internal) friction for each of the three regions.  These have been
identified by the straight line (red)
fits to those segments.
The most striking feature of the graph is
the peak which occurs at $36~s$.  Both to the left and to the right of
this peak the slope is the same $8.0~s^{-1}$.  At $68~s$, the slope
changes discontinuously
to the final segment value of $2.1~s^{-1}$.

These characteristics were found to be essentially
reproducible, as can be seen from figure 7; although the sharpness
of the peak is
more pronounced in figure 7.  If the STFT step-size had been 
smaller, to improve resolution; perhaps these would be more similar.
Although the peak in figure 7 occurs a little later
at $42~s$, it has roughly the same amplitude of $52~db$, corresponding to
an energy of $0.2~\mu J$.  Either side of the peak in figure 7 the slope
is the same $7.0~s^{-1}$, whereas the ending (third-segment) slope 
is $2.6~s^{-1}$.

\section{Energy Transition}
For Coulomb damping $\frac{d\Delta E}{dt}~=~R$ where $R$ is
a negative constant in the absence of
any change to the damping.  
The discontinuous change in the slope of Q that occurs
toward the end of the record, in each of figures 6 and 7, corresponds
to a downward jump in the rate at which energy is
lost per cycle.  If we designate the magnitude of this change in energy rate
by $|\delta(R)|$, then from the definition of Q one obtains
\begin{equation}
|\delta(R)|~=~\frac{2\pi E_0}{Q_0^{~2}}|\delta(\dot{Q})|
\end{equation}
\noindent where the subscript $0$ on $E$ and $Q$ identify
their values at the time of
the change.
From the level of the voltage in dB
at this point, and using the calibration constant of $48 V/rad$,
the amplitude was estimated at
$\theta_0~=~0.56~mrad$ in figure 6. Also at this transition, 
$Q_0~=~76$.
From eq(3) the energy is found to be \newline
$E_0~=~\frac{mg}{2}(r-a)\theta_0^{~2}$,
which yields $E_0~=~9.3\times 10^{-9}J$, upon substituting the
parameter values specified earlier. From the
slope change (estimated using trendline fits to each of the
segments), ~~
$|\delta(\dot{Q})|~=~5.6~s^{-1}$; and so finally, one obtains an estimate for
the energy jump in the rate of \newline
$|\delta(R)|~= ~5.7 \times 10^{-11} J/s$
Multiplying this rate by the period $T~=~0.45~s$ yields, for the energy
jump that occured at the transition
$\Delta\epsilon~=~2.6\times 10^{-11}~J.$
Repeating the process for figure 7 yielded the value $2.2\times 10^{-11}~J$.
Their average at
\begin{equation}
\Delta\epsilon_{est}~=~2.4\times 10^{-11}~J
\end{equation}
is noteworthy.

\section{Mesoscale Quantization}
The estimated energy transition indicated in eq(9) is
tantalizingly close to the
`hysteron' postulated by Erber and Peters\cite{01}, having the value
\begin{equation}
\Delta \epsilon_h~=~\frac{mc^2}{\alpha}~=~1.01\times 10^{-11} J
\end{equation}
\noindent where $mc^2$ is the rest energy of the electron and $\alpha$
is the fine structure constant.

\section{Self-organized Stability (SOS)}
A well-known fairly recent theory that applies to granular systems is 
`self-organized criticality' (SOC) \cite{03}.  Internal friction damping in
a mechanical oscillator is postulated to involve SOC.

The present author has postulated that creep, as with SOC, is at the
heart of internal friction damping of mechanical oscillators\cite{04}.
During creep,
structural members of an oscillator evolve toward metastable states.
Thus it appears that `self-organized stability' (SOS) is a process that is
likely to exist in the `sea' of SOC. The probability of an SOS
process being observable is expected to generally increase as the energy of
oscillation decreases.

It is postulated that the peaks in figures 6 and 7 are examples of
SOS.  Where the Q reaches its highest level, the energy decrement per cyle
has decreased to $|\Delta E| \approx 5 \times 10^{-10}J$.  Either
side of the peak the decrement is larger by roughly a factor of two.
Damping involves defect structures.  To decrease the energy decrement
is to decrease the number of defects operating collectively.  In other
words, to pass through a peak is to undergo a self-organized
stability
process that exists in transient form; i.e., a metastable state whose
lifetime we here quantify by the half-width of the SOS structure.  In figures
6 and 7 this is seen to be of the order of $5~s$.

\section{High-level Oscillation}
Reproducibility of SOS is observed by comparing figures 8 and 9, 
but there is no evidence for discrete
energy transitions as in figures 6 and 7. This is in keeping
with the earlier postulate that the transitions are more observable
at low energies.  Scatter (noise) in figures 8 and 9 is too large to
permit their observation. Observe also, that the half-life of the
SOS is roughly twice that of the lower energy cases; i.e., now about
$10~s$.

This author has held to the position for years now, that damping friction
is quantized at the mesoscale.  Evidence for his position has been
found in a variety of largely different experiments, the first of
which involved a low-frequency physical pendulum\cite{05}.  Another
example, involving a torsion pendulum, is described in an appendix of
\cite{01}.

It is assumed that dissipation of energy involves the removal of energy
`packets' from
the oscillator that are integer multiples of the hysteron energy
given in eq(10).  Whether the decay
is hysteretic (exponential) as in earlier experiments, or Coulomb in
character (present experiment); at larger energies of oscillation, the
discrete nature of the damping is not discoverable in the presence of
noise.  

\section{Surface Contamination}
The dramatic influence on damping of the surface state of cube and
cylinder is illustrated in figure 11.  Prior to the generation of
the free-decay record shown, both surfaces were touched with a finger
of the hand several
times and in several places, in the vicinity of the equilbirum
contact point. The fingerprints left on the surfaces resulted in
(i) a dramatic reduction in the Q, and (ii) a near perfect example of
Coulomb damping throughout the entire record. 

Similar records that are not shown, generated after applying various
fluids to the surfaces, also resulted in behavior similar to figure 11.
Of the three fluids considered for cleaning the surfaces:
(i) alcohol, (ii) a computer monitor cleanser, and (iii) acetone; the
monitor solvent of
unknown chemical composition caused enormous Q reduction, and
acetone resulted in the lowest dissipation.  The data of figures
3 \& 6 - 10 were all obtained after acetone cleaning, followed
by half a dozen decay runs before the start of data collection.  The
reason for the half-dozen pre-data `conditioning' is that the decay
curves were found to evolve significantly with time through
the first several runs following the
application of the acetone.

\clearpage

\begin{figure}
\includegraphics[width=4.0in]{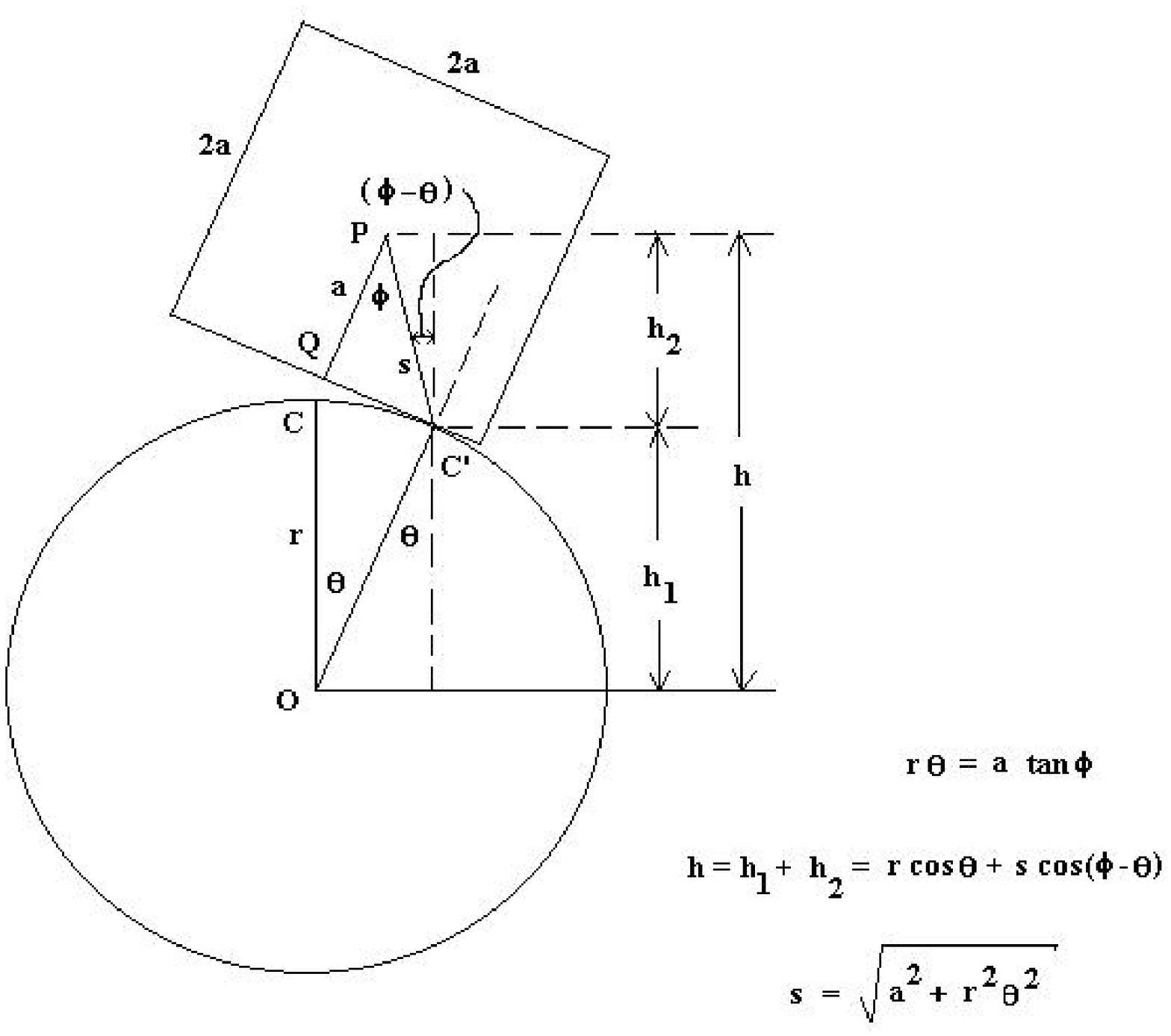}
\caption{\sloppy Illustration of the geometry of a
cube oscillating on a cylinder.}
\end{figure}

\begin{figure}
\includegraphics[width=4.0in]{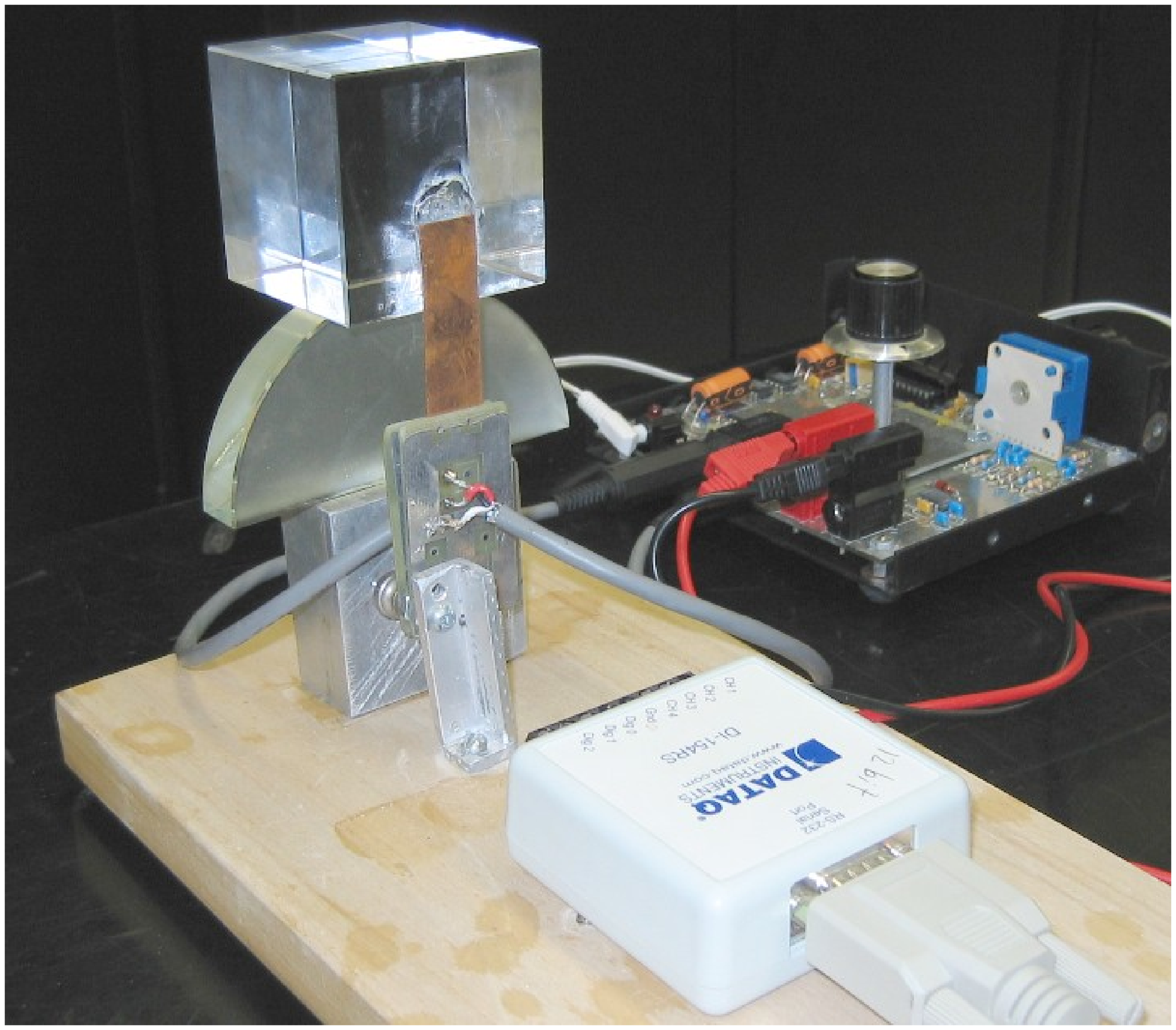}
\caption{\sloppy Photograph of the apparatus, showing oscillator,
sensor/electronics, and ADC.}
\end{figure}

\begin{figure}
\includegraphics[width=4.0in]{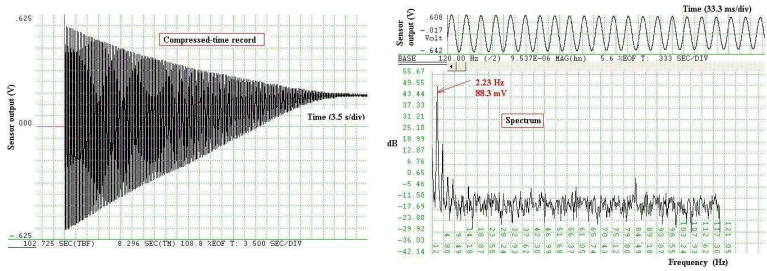}
\caption{\sloppy Low-level free-decay record and its spectrum.}
\end{figure}

\begin{figure}
\includegraphics[width=4.5in]{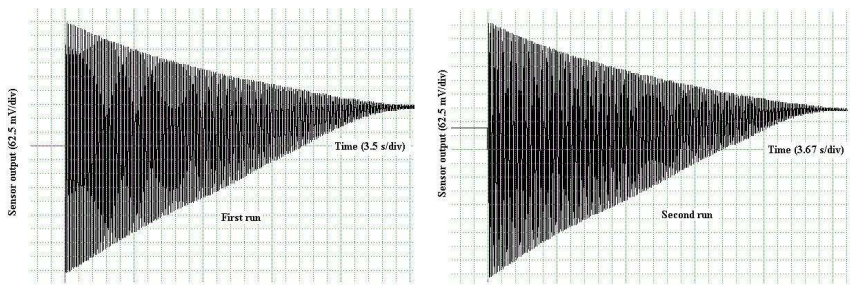}
\caption{Comparison of two low-level free-decay curves showing
a high level of reproducibility.}
\end{figure}

\begin{figure}
\includegraphics[width=4.5in]{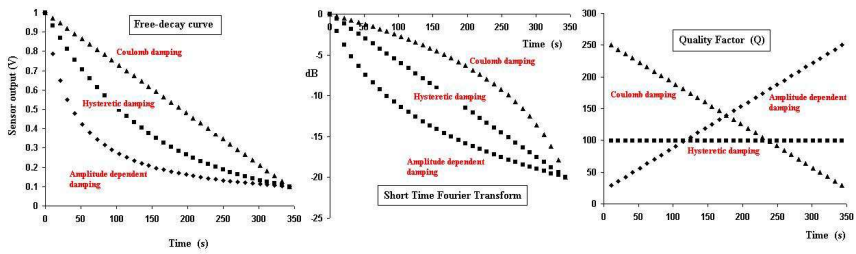}
\caption{Theoretical curves for three common forms of damping.}
\end{figure}

\begin{figure}
\includegraphics[width=4.5in]{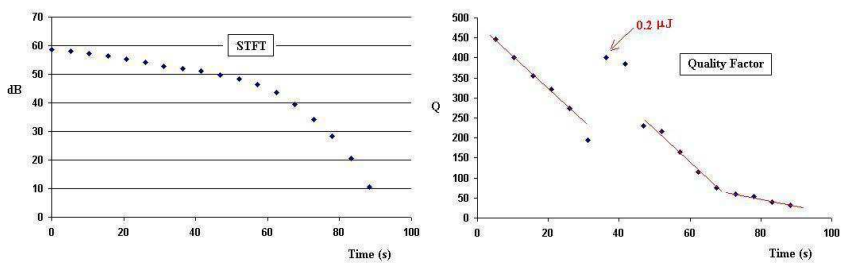}
\caption{STFT (left) from which the history of Q is obtained (right)
--corresponding to left graph of fig. 4.}
\end{figure}

\begin{figure}
\includegraphics[width=4.5in]{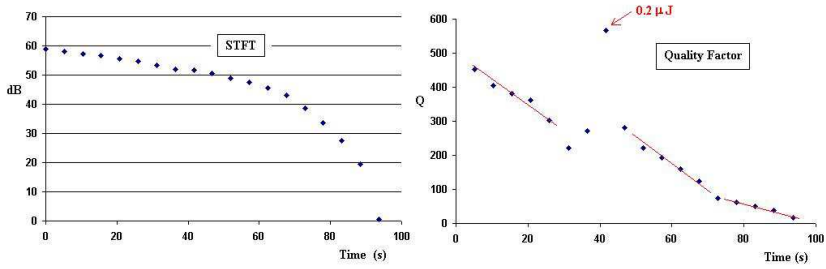}
\caption{Same as fig. 6, except corresponding to right graph of fig. 4.}
\end{figure}

\begin{figure}
\includegraphics[width=4.5in]{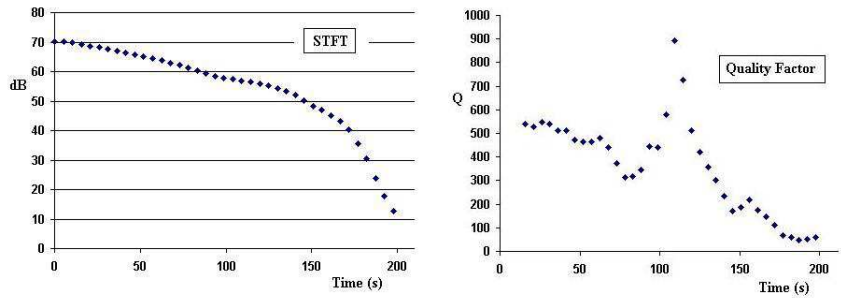}
\caption{\sloppy STFT and Q histories for a free-decay at higher starting
energy.}
\end{figure}

\begin{figure}
\includegraphics[width=4.5in]{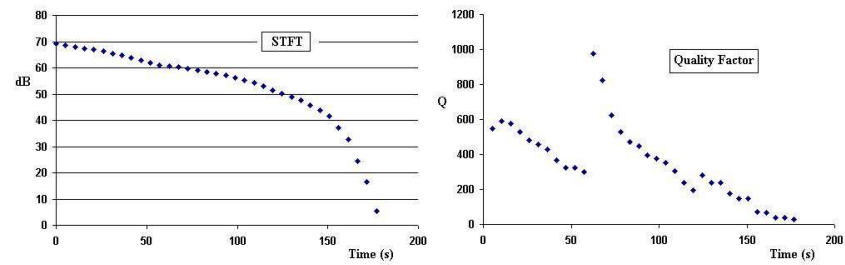}
\caption{\sloppy Repeat of figure 8 case, showing reproducibility of
SOS.}
\end{figure}

\begin{figure}
\includegraphics[width=4.5in]{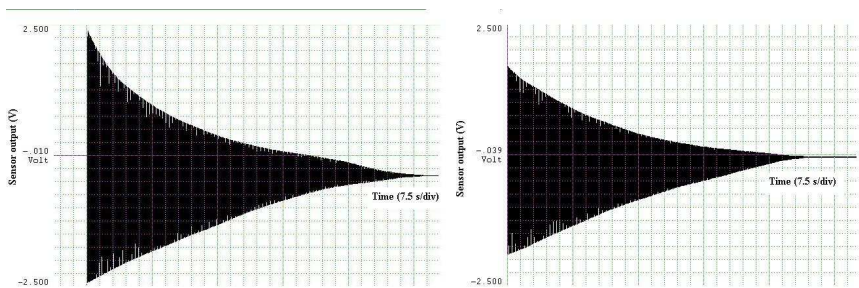}
\caption{\sloppy Free-decay records used to generate figures 8 (left)
and 9 (right).}
\end{figure}

\begin{figure}
\includegraphics[width=4.5in]{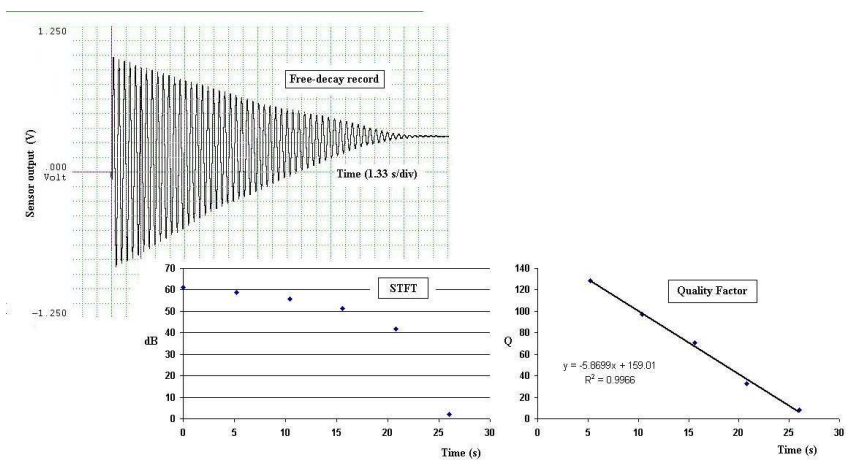}
\caption{\sloppy Illustration of the influence of surface contamination.}
\end{figure}

\end{document}